\documentclass{PoS}
\usepackage{wrapfig}

\title{Finite-size effects and the search for the critical endpoint in heavy ion collisions}

\ShortTitle{Finite-size effects and the search for the critical endpoint in heavy ion collisions}

\author{\speaker{Let\'\i cia F. Palhares}%
	\\
       Instituto de F\'\i sica, Universidade Federal do Rio de Janeiro, \\
Caixa Postal 68528, Rio de Janeiro, RJ 21941-972, Brazil\\
       E-mail: \email{leticia@if.ufrj.br}}

\author{Eduardo S. Fraga\\
        Instituto de F\'\i sica, Universidade Federal do Rio de Janeiro, \\
Caixa Postal 68528, Rio de Janeiro, RJ 21941-972, Brazil\\
        E-mail: \email{fraga@if.ufrj.br}}
	
\author{Takeshi Kodama\\
        Instituto de F\'\i sica, Universidade Federal do Rio de Janeiro, \\
Caixa Postal 68528, Rio de Janeiro, RJ 21941-972, Brazil\\
        E-mail: \email{tkodama@if.ufrj.br}}	

\abstract{We discuss how the finiteness of the system created in a heavy-ion collision affects possible signatures of the QCD critical endpoint. We show sizable results for the modifications of the chiral phase diagram at volume scales typically encountered in current heavy-ion collisions and address the applicability of finite-size scaling as a tool in the experimental search for the critical endpoint.}

\FullConference{5th International Workshop on Critical Point and Onset of Deconfinement\\
                June 8-12, 2009\\
                Brookhaven National Laboratory, Long Island, New York, USA}

\begin{document}

\section{Introduction}

The existence and the position of the chiral (second-order) critical endpoint (CEP) are key questions in mapping the phase diagram of strong interactions. Recently, the possibility that heavy-ion collisions (HICs) may probe experimentally this point has generated an increasing interest in how it would affect different observables, yielding possible signatures.

Since the CEP is associated with a second-order phase transition and diverging correlation length,
some of the signatures \cite{Stephanov:1998dy} are based on the expected (divergent) critical behavior of the
correlation functions of the quasi-particle $\sigma$, related to the order parameter of  
the chiral transition:
\begin{equation}
\langle\sigma^n \rangle \sim \xi^{p_n} \, ,
\end{equation}
where $\xi$ is the correlation length and $p_n$ is a positive exponent.
This feature should be translated into final observables in a HIC via mesonic decays of the sigma field into other particles, especially {\it soft} pions (created as soon as the medium-dependent $\sigma$ reaches the mass threshold).

\begin{wrapfigure}{R}{0.45\textwidth}
%
%
\vspace{-.8cm}
  \begin{center}
    \includegraphics[width=.45\textwidth]{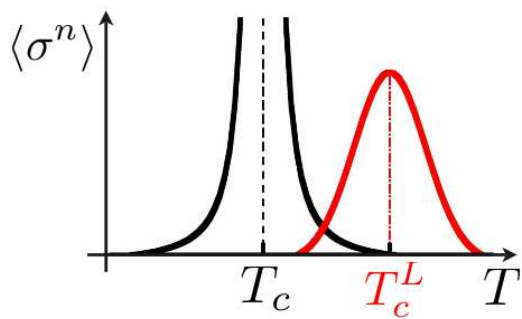}
  \end{center}
  \caption{The $n$-point correlation function of the order parameter $\sigma$ is shown as a function of the external parameter $T$. The singular (black) curve corresponds to the thermodynamic limit, while the behavior in a finite system is illustrated by the shifted peak (red curve). }\label{pseudocorr}
%
%
\end{wrapfigure}

In any real system, however, the correlation length is trivially bounded by the finite volume and, through causality\footnote{In principle, one has to consider the growth rate of correlated domains in a fully dynamical approach, the speed of light (i.e. causality) being the ultimate bound in a relativistic system.}, by the finite lifetime of the system. Instead of the divergent critical behavior expected for correlation functions in the thermodynamic limit, the observables measured will show {\it pseudocritical} peaks, corresponding to a smoothening of the associated critical singularity. As illustrated in Figure \ref{pseudocorr}, these peaks are actually shifted from the position of the original singularity by a size-dependent amount.

In HIC experiments, the volume of the system is finite and centrality-dependent \cite{Adams:2005dq}. The typical linear sizes $L$ are actually quite small ($L\lesssim d_{\rm ion}$, with the ion diameter $d_{\rm ion} \approx 10-15~{\rm fm}$), corresponding to energies $L^{-1}\gtrsim 13.1-19.7~ {\rm MeV}$, not negligible in comparison with the expected scales in the critical region of the phase diagram $T_c\lesssim 200~{\rm MeV}$, especially in non-central collisions. This suggests that finite-volume effects may affect significantly the different physical phenomena occuring in a heavy-ion collision.

Nevertheless, these effects have been mostly overlooked in the descriptions of phenomena related to the quark-gluon plasma, with some exceptions
 (e.g. Refs. 
 \cite{Gopie:1998qn,Spieles:1997ab,Fraga:2003mu,Braun:2004yk,finite-NJL,Yamamoto:2009ey,Elze:1986db,Bazavov:2007zz}).

Here, we are concerned with the role played by finite-size effects on the QCD phase diagram and on the physics of the CEP. In particular, we address the following question: how can the centrality-dependent, finite volume of the system created in HICs affect possible signatures of the CEP?
Using two different approaches, we point out that finite-size effects are relevant and should be thoroughly investigated in the context of the QCD phase diagram probed in HICs and may play an important role in the experimental search for the CEP. In Section \ref{Shifts}, we show estimates of the amplitudes of the shifts of pseudocritical lines in the chiral phase diagram at typical HIC size scales. The large effects obtained suggest that the actual phase diagram probed in HICs can be quantitatively very different from the usual picture in the thermodynamic limit. In Section \ref{FSS}, we discuss how one can take advantage of the fact that HIC data are constituted of a set of systems with different volumes to construct a complementary tool for the experimental search of the CEP. Finally, we conclude and discuss some perspectives in Section \ref{Conclusions}.

\section{Volume dependence of the chiral CEP at HIC scales}\label{Shifts}

To investigate whether the pseudocritical phase diagram probed in heavy-ion collisions differs significantly from the usually adopted picture in the thermodynamic limit, we estimated \cite{Palhares:2009tf} the shift of the critical line in the temperature-chemical potential plane in a well-established
\cite{quarks-chiral,Scavenius:2000qd,Scavenius:2001bb}, chiral effective model: the linear sigma model with constituent quarks \cite{GellMann:1960np}, described by the Lagrangian density
\begin{equation}
\mathcal{L} = \overline{\psi}_{f} \left[ i\gamma^{\mu}\partial_{\mu} +
\mu\gamma^{0} - g\sigma)\right] \psi_{f} + \frac{1}{2}\partial_{\mu}\sigma%
\partial^{\mu}\sigma- V(\sigma)\;,  \label{lagrangian}
\end{equation}
where $\mu$ is the quark chemical potential, 
\begin{equation}
V(\sigma)=\frac{\lambda}{4}(\sigma^{2} - \mathit{v}%
^{2})^{2}-h\sigma  \label{bare_potential}
\end{equation}
is the self-interaction potential for the mesons, exhibiting both
spontaneous and explicit breaking of chiral symmetry. The pions were dropped for simplicity and all parameters are fixed to reproduce known properties (observed or simulated on the lattice) of the QCD vacuum (cf. Ref. \cite{Palhares:2009tf} and references therein for details). As usual, the scalar field $\sigma$ plays the role of an approximate order parameter for the chiral transition, being an exact order parameter for massless quarks. Investigating how this field changes with temperature and chemical potential we obtain the phase diagram for the chiral effective model.

Our main goal here is to establish a well-defined and clean measure of the finite-size corrections to the position of the chiral CEP, so that we can investigate their relevance in the context of HICs. In this vein, we concentrate on bulk effects, restricting our analysis to the comparison within a simple (though largely adopted) picture: the {\it equilibrium} mean-field approximation\footnote{The results for the thermodynamic limit have been discussed in Ref. \cite{Scavenius:2000qd}}, neglecting completely fluctuations and inhomogeneities. 

Therefore, we calculate the effective potential of the model in the mean-field approximation, including quark thermal fluctuations to one loop.
In the case of a finite system of linear size $L$, the momentum integral
from the one-loop quark contribution to the effective potential is
substituted by a sum 
\begin{equation}
\frac{V_{q}}{T^{4}}=\frac{2N_{f}N_{c}}{(LT)^{3}}\sum_{\mathbf{k}}\left[ \log
\left( 1+e^{-(E_{\mathbf{k}}-\mu )/T}\right) +\log \left( 1+e^{-(E_{\mathbf{k%
}}+\mu )/T}\right) \right] \;,
\end{equation}%
where $E_{\mathbf{k}}=\sqrt{\mathbf{k}^{2}+m_{eff}^{2}}$, and $%
m_{eff}=g|\sigma |$ is the effective mass of the quarks.
The boundary conditions enter the expression above via the dispersion relation $\mathbf{k}(\mathbf{n})$
with $n_i$ being the labels of the discrete spatial Fourier modes. The sensitiveness of the 
results to different boundary conditions is then illustrated by the comparison between two extreme
cases, periodic (PBC; $k_i=2\pi n_i$) and antiperiodic (APC; $k_i=(2n_i+1)\pi$) boundary conditions.

\hspace{.6cm}
\begin{figure}[ht!]
\begin{minipage}{.5\linewidth}
\includegraphics[width=\linewidth]{CEP}
\caption{Trajectories of the pseudo-CEP as the system size is decreased for PBC (dotted line) and APC (dashed line). }\label{CEP}
\end{minipage}
\hspace{.5cm}\vspace{1cm}
\begin{minipage}{.46\linewidth}
    \includegraphics[width=\linewidth]{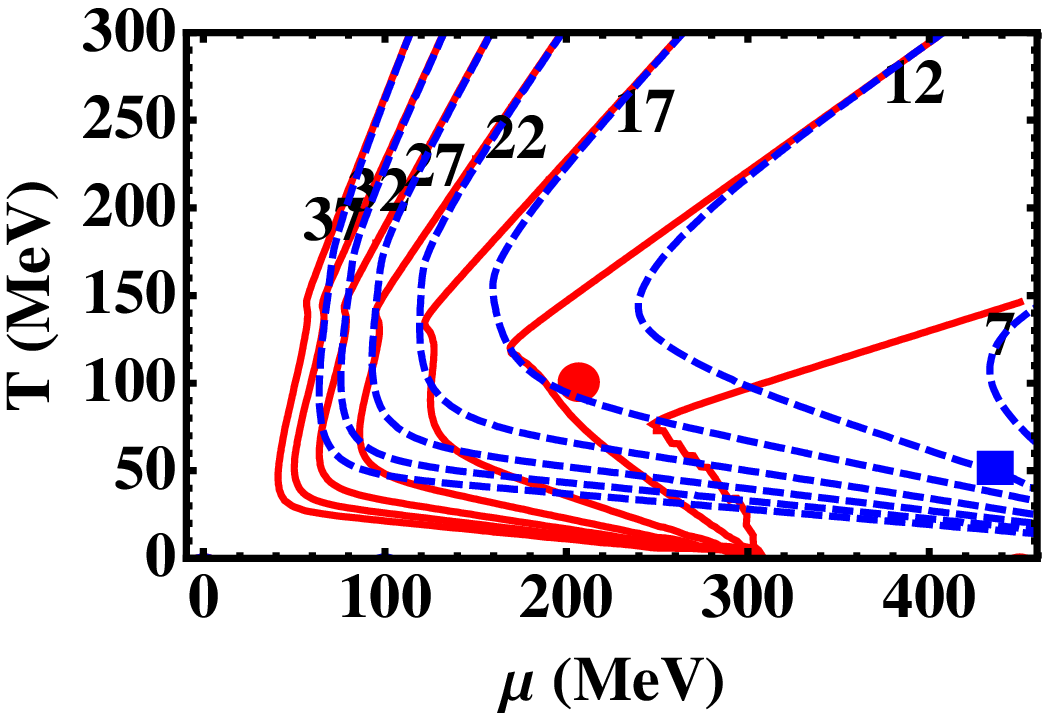}
\caption{Isentropic lines in the $T-\mu$ plane in the thermodynamic limit (red, solid line) and for a finite system with $L=2~ $fm (blue, dashed line), labeled by the value of entropy per baryon number $s/n_B$.}\label{Isentropics}
\end{minipage}
\end{figure}

\vspace{-.5cm}
\begin{wrapfigure}{r}{0.45\textwidth}
%
%
\vspace{-.6cm}
  \begin{center}
  \includegraphics[width=\linewidth]{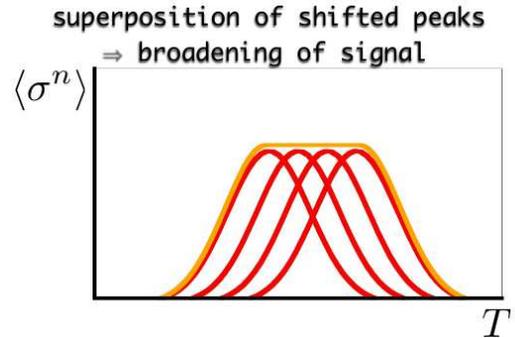}
  \end{center}
\vspace{-.4cm}
\caption{Cartoon illustrating the broadening of a pseudocritical signal due to the averaging over a centrality interval.}\label{broadening}
%
%
\end{wrapfigure}

One can calculate \cite{Palhares:2009tf} the transition lines representing pseudo-first-order transitions, characterized
by a discontinuity in the approximate order parameter, the chiral condensate 
$\sigma$, and the production of latent heat through the process of phase
conversion. The CEP is then identified as the end of those lines. We find that those lines are displaced to the region of higher $%
\mu$ and shrinked by finite-size corrections. The former effect is sensibly
larger when PBC are considered, indicating that the presence of the spatial
zero mode tends to shift the transition region to the regime of larger
chemical potentials. Both boundary conditions reproduce the infinite-volume
limit for $L\gtrsim10$ fm. Figure \ref{CEP} shows the corresponding
displacement of the pseudocritical endpoint, comparing PBC and APC: both
coordinates of the critical point are significantly modified, and $\mu_{%
\mathrm{CEP}}$ is about $30\%$ larger for PBC. Our findings for the 
corresponding isentropic trajectories in the $T-\mu$ plane are 
shown in Figure \ref{Isentropics},
comparing the infinite-volume limit with the finite system with $L=2$ fm
in the case with APC. For sufficiently high temperatures, the
isentropic lines in the thermodynamic limit are reproduced, while large
discrepancies are found around and below the transition region.
The reader is referred to Ref. \cite{Palhares:2009tf}
for further details and plots.

Our results clearly indicate that finite-size corrections to the chiral phase diagram at volume scales typically encountered in HICs can be large and therefore different signatures of the CEP should be affected. First of all, signatures related to the nonmonotonic behavior of fluctuation measurements are actually probing the pseudocritical peak that would occur at significantly larger $\mu$ as compared to the usual expectation in the thermodynamic limit. Figure \ref{Isentropics} indicates that signatures based on the focussing of isentropic trajectories could probe a quantitatively different scenario due to the finiteness of the systems created at HICs. It is also interesting to note that, since the system size realized at HICs depends on the centrality of the collision, measurements obtained as averages over not sufficiently small centrality windows could broaden the already smoothened pseudocritical peaks (as illustrated in Figure \ref{broadening}), helping to hide it in the background. Of course, the width of the centrality bins in data analysis is also bounded from below so that the statistics is enough to guarantee reasonable statistical errors.


\section{Finite-size analysis as a tool for searching the CEP in HICs}\label{FSS}

Now that we have argued that finite-size corrections to critical phenomena in the context of HICs can be sizable, we can approach this issue from a different perspective with the aim of taking advantage of this feature in the experimental search for the CEP.

As is well-known \cite{DL,fisher,amit}, the second-order transition that occurs at the CEP in the thermodynamic limit is characterized by a divergent correlation length and by the property of scale invariance. These constraints in the thermodynamic limit are translated into real, finite systems as the existence of finite-size scaling (FSS) \cite{Cardy:1996xt,Brezin:1981gm,Brezin:1985xx} in the vicinity of criticality. The phenomenon of FSS can be rigorously proved through a renormalization group analysis and is extensively studied and successful in condensed matter physics (cf. Ref. \cite{sldq} for an example within the context of spin glass transitions in disordered Ising systems).

\begin{figure}[h!]
\begin{minipage}{.55\linewidth}
\includegraphics[width=.85\linewidth]{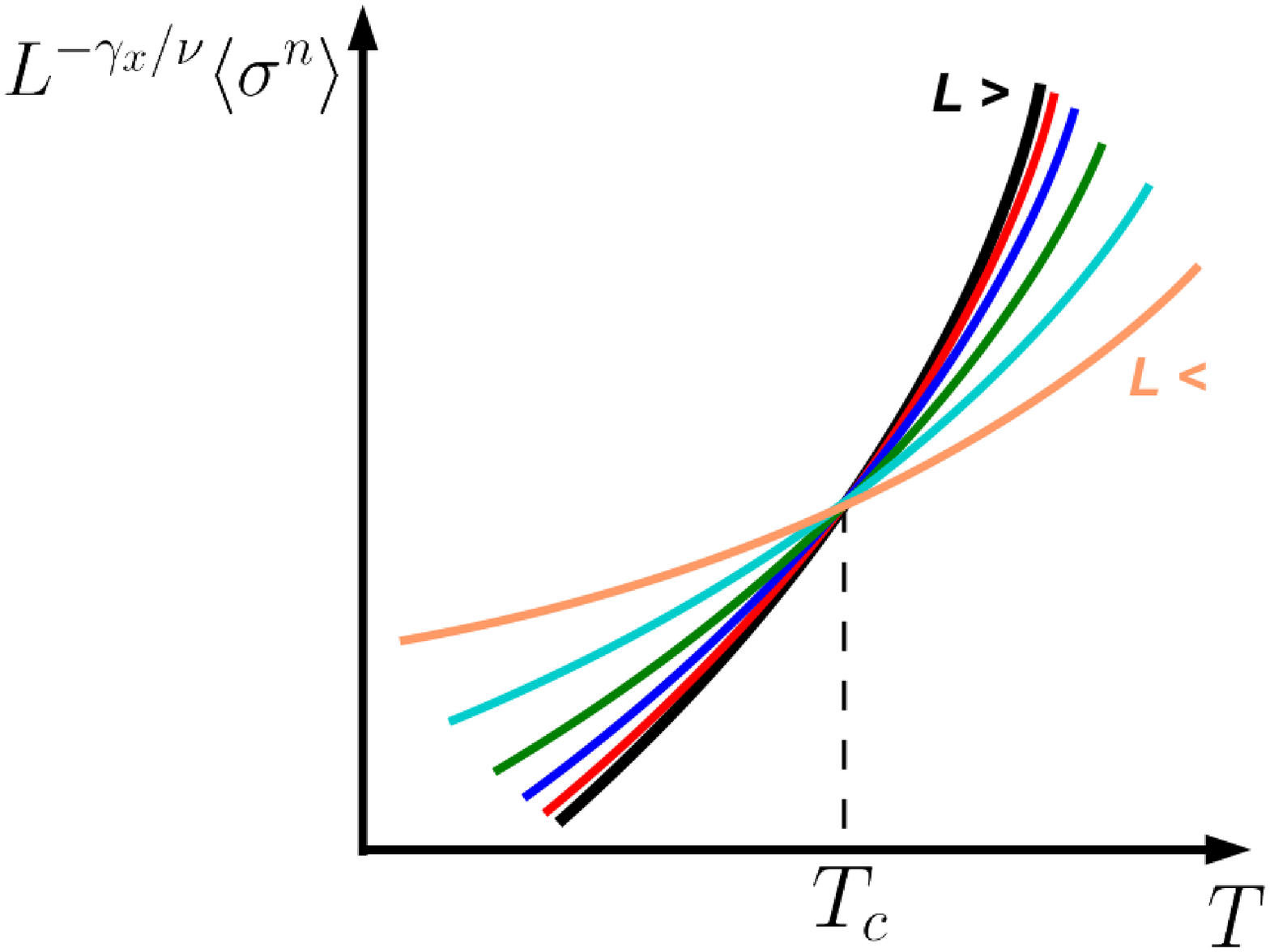}
\end{minipage}
\hspace{-.4cm}
\begin{minipage}{.55\linewidth}
\includegraphics[width=.9\linewidth]{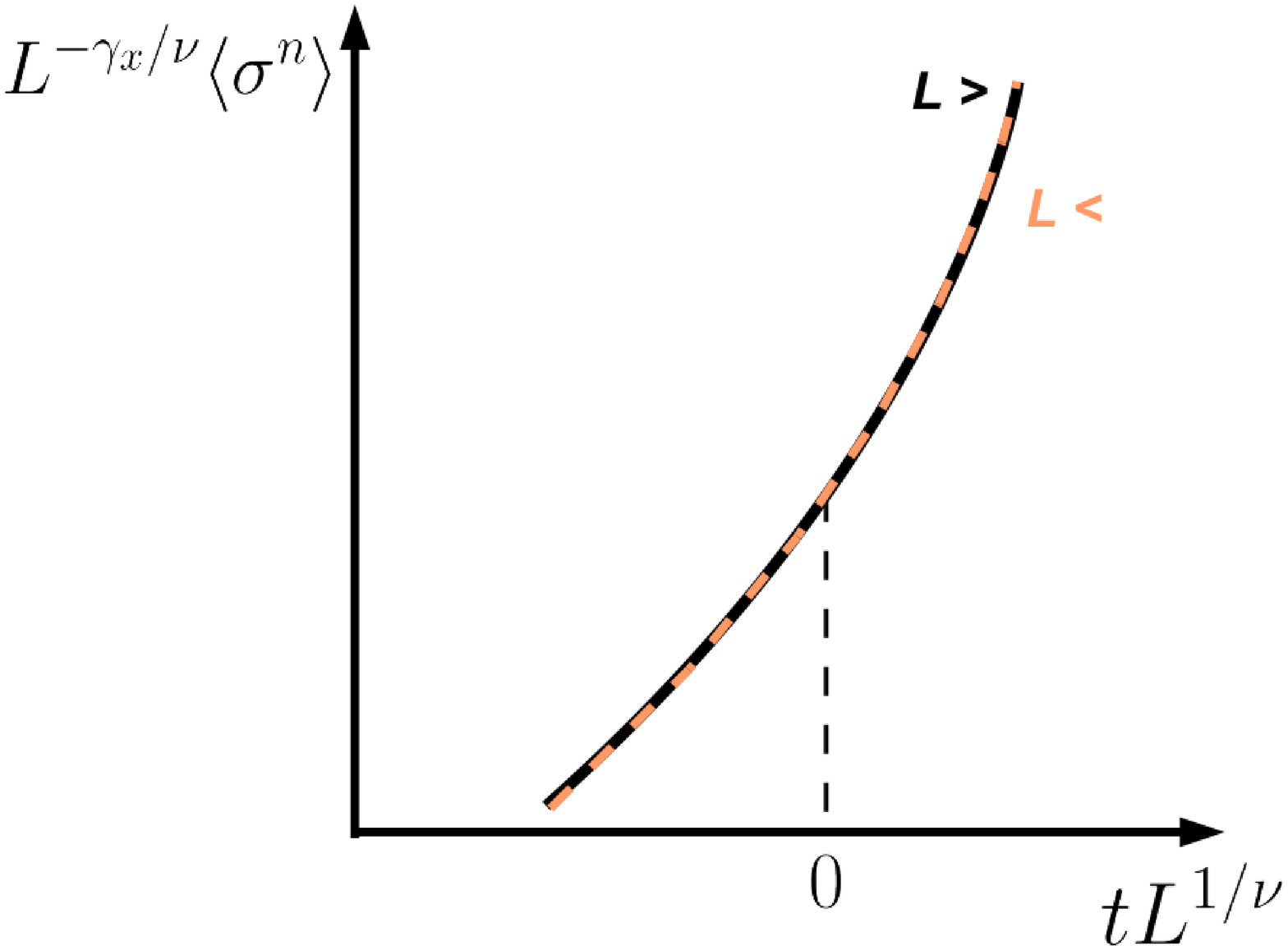}
\end{minipage}
\caption{On the left, the normalized correlation function is shown as a function of external parameter $T$, with different curves corresponding to different system sizes. The plot on the right illustrates the collapsing of curves in the appropriate scaling plot.}\label{collapse}
\end{figure}

One can state the FSS via its consequences on the correlation functions of the order parameter. In the FSS regime, any correlation function $X(T,L)$ of the order parameter does not depend independently on the external parameter $T$ and on the size $L$ of the system, having the following scaling form \cite{Cardy:1996xt}:
\begin{equation}
X(T,L)=L^{\gamma_x/\nu}f_x(tL^{1/\nu})\, ,\label{scaling}
\end{equation}
where $t=(T-T_c)/T_c$ represents a dimensionless measure of the distance, in the external parameter domain, to the genuine CEP (in the thermodynamic limit), $\gamma_x$ is a dimension exponent and $\nu$ is the universal critical exponent defined by the divergence of the correlation length. The scaling form in Eq. (\ref{scaling}) implies (and is implied by)\footnote{One can consider the case in which the system is exactly {\it on} the CEP (cf. the talk by W. Yuanfang in this conference and Ref. \cite{Lizhu:2009md}), then $t=0$ and the scaling relation Eq. (\ref{scaling}) reduces to $X\propto L^{\gamma_s/\nu}$. This is a necessary {\it but not sufficient} condition for the presence of FSS in a given data set and, therefore, it does not imply the presence of a CEP.} the existence of a {\it scaling plot} in which all the curves for different system sizes collapse into a single curve, as illustrated in Figure \ref{collapse}. 

\hspace{-.9cm}

Therefore, a systematic procedure to search for a CEP is to obtain an ensemble of data with nontrivial size dependence for a given correlation function $X$ and look for $T_c$, $\gamma_x$ and $\nu$ such that the size dependence disappears in the appropriate scaling plot. The same values of $T_c$ and $\nu$ should then generate collapsed curves for all observables directly connected to any correlation function $X$; the only free parameter in this second stage being the dimension exponent $\gamma_x$. If the scaling is found, then one not only concludes that there {\it exists} a CEP in the vicinity of the external parameter domain of the analyzed data, but also determines, in the process, the location of the genuine CEP (in the thermodynamic limit) and its critical exponents.

 Another interesting feature of the FSS as a tool for searching the CEP is that one can test for the FSS hypothesis even if the data set is restricted to regions only above (or below) the CEP, provided the system is in the vicinity of the criticality.
In a realistic case, the collapsing of curves should be realized within experimental errors and a careful statistical treatment should be implemented. 

The data generated in HICs indeed correspond to measurements from an ensemble of systems of different sizes, so that in principle one can implement a FSS analysis in this case.
In order to apply FSS as a tool for searching the CEP in HICs, it is necessary to
translate the scaling relation in Eq. (\ref{scaling}) to the physics scenario encountered in HICs. Since the scaling relation will hold for any variables $t'$ and $L'$ directly proportional to the real dimensionless distance $t$ to the CEP and size $L$ of the system, respectively, we need to determine experimental observables only up to normalization constants.There also exist scaling relations for variables proportional to powers of $t$ and $L$; the only difference in this case is that the parameters $\gamma_x$ and $\nu$ obtained in the FSS analysis will not be the genuine critical exponents of the theory.

Let us now translate the scaling relation, Eq. (\ref{scaling}), to the HICs scenario.
Firstly, the observables that satisfy FSS relations are those directly connected to the different correlation functions of the chiral order parameter, that should be related to fluctuations in the final spectra at HICs via mesonic decays. The size of the system $L$ is taken to be proportional to the root of the number of participants, due to the approximately two-dimensional initial condition. Finally, describing the distance to the CEP in HICs is a more subtle issue, because there are two important external parameters\footnote{Actually, there are also two different critical exponents $\nu_T$ and $\nu_{\mu}$, since the divergence of the correlation length will be different in these two directions, as it happens for the Ising model with temperature and external magnetic field.}: the temperature $T$ and the chemical potential $\mu$. Determining (or even defining, in the case of a dynamics far from equilibrium) the direction in which the system
created in a HIC approches a given point in the phase diagram is an extremely difficult task. Fortunately, the implementation of a FSS investigation does not require this information, but rather a reasonable measure of the distance between the CEP and the external parameters that characterize the data to be analyzed.
As suggested by the success of thermal statistical models, the final measured particle spectra should reflect the {\it freeze-out} values of $T$ and $\mu$, that are constrained to the freeze-out curve in the $T-\mu$ plane and, therefore, do not represent totally independent parameters. Since the freeze-out curve is naturally parameterized by the center-of-mass energy $\sqrt s$ of the collision, one simple and reasonable prescription is to define the distance, in the external parameter space, to the CEP via $\delta s\equiv (\sqrt s -\sqrt{s_c})/\sqrt{s_c}$, where $\sqrt{s_c}$ is the center-of-mass energy corresponding to the genuine CEP. Finally, we arrive at the scaling relation for the case of HICs:
\begin{equation}
X~N_{\rm part}^{-\gamma_x/2\nu}=f_x\left( \frac{\sqrt{s}-\sqrt{s_c}}{\sqrt{s_c}}~N_{\rm part}^{1/2\nu} \right)\,. \label{scalingHIC}
\end{equation}
In practice, the search for FSS in HIC data is equivalent to verifying if the scaling relation in Eq. (\ref{scalingHIC}) is compatible with the measured observables. This can be implemented via conveniently constructed $\chi^2$ methods (as the one proposed in Ref. \cite{Palhares:2009tf}), aiming at the minimization of the distances in the scaling plot in such a way that the curves collapse into a single one.

Of course, there are many caveats to be overcome and kept in mind in applying this technique to HIC data. As most of the proposed signatures for the CEP in HICs, the applicability of a FSS analysis to the measurements relies on the fact that the critical correlations are not completely washed out by the evolution in the hadronic phase nor hidden within errors due to the large thermal background. Nevertheless, it represents a complementary and independent tool for the search of the CEP that may help constructing a robust and consistent interpretation of the critical behavior of the strongly interacting matter generated in current and in future heavy-ion collider experiments.

Finally, as stated above, we have completely disregarded dynamical phenomena in the discussion of finite-size effects. In an out-of-equilibrium approach, one has to consider the fact that it takes a finite time for the correlation length to grow, even on the CEP. If the lifetime of the system is a more constraining bound than its actual size, the correlated domains will not grow as large as the system and this original formulation of the FSS analysis would not be applicable. However, in heavy-ion collisions the lifetime of the system is also an energy- and centrality-dependent quantity and, in this case, one alternative idea would be to try to simply extend this analysis using a ``horizon'' size (defined via the lifetime and the sound velocity, or ultimately the speed of light) instead of the size of the system. 

\section{Final remarks}\label{Conclusions}

In these proceedings, we have complemented the discussion (originally addressed in Ref. \cite{Palhares:2009tf}) on different aspects of finite-size effects that render crucial features in various proposed signatures of the QCD critical endpoint in HICs. We have discussed how the phase structure and the critical behavior of the strongly interacting matter created in HICs should differ from the usual picture in the thermodynamic limit. 
Using two different approaches, we argue that the finiteness of the system created in HICs may play an important role in the search for the CEP, especially at the Beam Energy Scan program \cite{BES} to start soon at RHIC-BNL.

In particular, we have estimated the amplitude of the shifts of the critical line and the position of the CEP in the chiral phase diagram at HIC size scales and shown that the modifications can be large and
generate sizable effects in different signature scenarios for the CEP. We have further discussed the implementation of FSS analysis as a complementary tool for the search of the CEP, taking advantage of the fact that HIC data are actually measurements from an ensemble of systems of different sizes.

Finally, we believe that the results above strongly suggest that, in the context of HICs, the finiteness of the system is a crucial feature whose consequences should be thoroughly investigated in the different phenomena characterizing the new state of matter probably created in these experiments.

\section*{Acknowledgements}
The authors thank the organizers of the 5th International Workshop on Critical Point and Onset of Deconfinement for providing such a nice environment for the debate of ideas. We are grateful to A. Mocsy, K. Rajagopal, K. Redlich, M. Stephanov, and especially to P. Sorensen for stimulating discussions. This work was partially supported by CAPES, CNPq, FAPERJ and FUJB/UFRJ.

\end{document}